\documentclass[fleqn,10pt]{elsart}
\usepackage{amssymb}
\usepackage{indentfirst}
\usepackage{psfig,color}
\usepackage{epsfig}
\usepackage{epsf}
\usepackage{graphicx}

\journal{Nuclear Physics A}

\begin{document}

\begin{frontmatter}
\title{Sivers function in light-cone quark model and
azimuthal spin asymmetries in pion electroproduction}

\author[pku]{Zhun L\"{u}},
\author[ccastpku]{Bo-Qiang Ma\corauthref{cor}}
\corauth[cor]{Corresponding author.} \ead{mabq@phy.pku.edu.cn}
\address[pku]{Department of Physics, Peking University, Beijing 100871, China}
\address[ccastpku]{CCAST (World Laboratory), P.O.~Box 8730, Beijing
100080, China\\
Department of Physics, Peking University, Beijing 100871, China}

\begin{abstract}
We perform a calculation of Sivers function in a light-cone SU(6)
quark-diquark model with both scalar diquark and vector diquark
spectators. We derive the transverse momentum dependent light-cone
wave function of the proton by taking into account the
Melosh-Wigner rotation. By adopting one-gluon exchange, we obtain
a non-vanishing Sivers function of $down$ quark from interference
of proton spin amplitudes. We analyze the $\frac{|P_{h\perp}|}{M}$
weighted Sivers asymmetries in $\pi^+$, $\pi^-$ and $\pi^0$
electroproduction off transverse polarized proton target, averaged
and not averaged by the kinematics of HERMES experiment.
\end{abstract}

\end{frontmatter}

\section{Introduction}
The investigation of single-spin asymmetries in hadronic reaction
has attracted great attention because it can uncover some novel
structure of the nucleon. Various theoretical explanations have
been put forward. One of them is the so-called Sivers effect
\cite{sivers}, which indicates that the transverse motion of
quarks in transverse polarized proton can lead to an azimuthal
production asymmetry. The effect can be factorized as a
convolution of a T-odd transverse momentum dependent parton
distribution function (Sivers function)
$f_{1T}^{\perp}(x,\mathbf{k}_\perp)$, the number density of
unpolarized quarks in transversely polarized proton, with the
unpolarized fragmentation function. However, for several years the
Sivers function was believed to vanish due to the time reversal
invariance property of QCD. An alternative mechanism hereafter
named the Collins effect was proposed in Ref.~\cite{ref:col1}: the
chiral-odd fragmentation function convoluted with the transversity
distributions of the proton can generate the single-spin
production asymmetries. Collins' idea motivated the experimental
and theoretical efforts to reveal the transversity properties of
the nucleon~\cite{ref:barone}. Semi-inclusive DIS (SIDIS)
experiments on longitudinally polarized target by HERMES
collaboration~\cite{ref:hermes1} and on transversely polarized
target by SMC collaboration~\cite{ref:smc} showed significant
azimuthal asymmetries. During the same time, a number of
theoretical analyses on these observed asymmetries have been
undergone in terms of the Collins
effect~\cite{ref:am,ref:msy,ref:efremov,ref:oganessyan}. Recently,
Brodsky, Hwang and Schmidt~\cite{ref:bhs,ref:bhs2} proposed a new
mechanism of producing single spin asymmetries, and they
demonstrated that the final state interactions from gluon exchange
between the outgoing quark and target spectator system can lead to
single-spin asymmetries in SIDIS or Dell-Yan process at
leading-twist level in perturbative QCD. It was shown soon by
Collins~\cite{ref:col2} that this mechanism can be recognized as
Sivers effect, and the Sivers function is not vanish due to the
presence of Wilson line in the operator definitions of parton
density. This implies that the gauge link term in the
gauge-invariant definition of the parton distribution function can
generate the final or initial state interactions~\cite{ref:ji},
which are necessary to produce a phase difference for a nonzero
Sivers function.

The allowance of T-odd parton distribution opens up a wide range
of phenomenological applications. In this work we attempt to give
a full analysis of Sivers asymmetries in SIDIS in the scope of
final state interactions. Based on the quark-scalar-diquark model
employed in~\cite{ref:bhs}, we incorporate vector diquark
structure into the proton wave function through a simple
relativistic quark spectator-diquark model which is formulated in
the light-cone frame. This model was originally proposed in order
to study deep inelastic lepton nucleon
scattering~\cite{feynman,close}, based on the quark-parton model
picture that deep inelastic scattering is well described by the
impulse approximation, in which the incident lepton scatters
incoherently off a quark in the nucleon, with the remaining
nucleon constituents treated as a quasiparticle spectator. After
taking into account Melosh rotation effects, this model is in good
agreement with the experimental data of polarized deep inelastic
scattering, and the mass difference between the scalar and vector
spectators reproduces the $up$ and $down$ valence quark
asymmetry~\cite{ma}. By advantage of the proton wave function we
find that the interference of the proton spin amplitudes can
provide necessary phase for Sivers function of $down$ quark. The
inclusion of vector diquark has also been adopted in
Ref.~\cite{ref:bsy}, where a version of the axial-vector spectator
model~\cite{jmr} was employed. Although there are vector diquarks
in both models, we will explain later that, our model is different
from that in Ref.~\cite{ref:bsy}. Both models can produce
qualitatively similar Sivers functions and asymmetries, but our
model has fewer parameters. Relying on Sivers functions of $up$
and $down$ quarks we give some prediction for the Sivers
asymmetries in $\pi^+$ production in SIDIS, and also rough
estimation of those in $\pi^-$ and $\pi^0$ production which is
currently analyzed in HERMES Collaboration. One feature of our
model which will be shown is that if we take away the vector
diquark spectator, the result of our model returns to the result
appearing in Ref.~\cite{ref:bhs}.

This paper is organized as follows. In Sec.~2 we derive the
light-cone wave function for the two-body Fock state of the proton
including vector diquark spectator utilizing the Melosh-Wigner
rotation. Then we estimate the Sivers functions of $up$ and $down$
quarks in the presence of final state interactions in Sec.~3. In
Sec.~4, we analyze the single-spin Sivers asymmetries of $\pi^+$,
$\pi^-$ and $\pi^0$ off proton target respectively. Finally we
give a brief summary in Sec.~5.

\section{The quark-diquark decomposition of proton in Fock space}
The light-cone formalism~\cite{PinskyPauli} has remarkable
properties in describing composite state such as hadron. It is
suitable to describe the relativistic many-body problem and there
have been many successful applications of the light-cone quark
model to various physical processes~\cite{lcap}. In light-cone
formalism, the hadronic wave function is expressed in terms of a
series of light-cone wave functions in the Fock state basis. For
the proton a convenient basis is the quark-diquark two-body state,
which has been adopted in Ref.~\cite{ref:bhs}, where the two-body
state wave function was derived from the relativistic field theory
treatment by calculating the interaction vertex in light-cone
frame~\cite{ref:bhs}:
\begin{equation}
 \frac{\bar{u}(k^+,k^-,\mathbf{k}_\perp)}{\sqrt{k^+}}\cdot
 \frac{u(P^+,P^-,\mathbf{P}_\perp=\mathbf{0}_\perp)}{\sqrt{P^+}},
\end{equation}
$\bar{u}(k^+,k^-,\mathbf{k}_\perp)$ and
$u(P^+,P^-,\mathbf{P}_\perp)$ are the light-cone spinors of the
quark and the proton respectively. A similar treatment was
employed in Refs.~\cite{BD80,ref:bhms} where the electron-photon
two-particle Fock state decomposition was obtained from the
interaction vertex:
\begin{equation}
 \frac{\bar{u}(k^+,k^-,\mathbf{k}_\perp)}{\sqrt{k^+}}\gamma\cdot\epsilon^*
 \frac{u(P^+,P^-,\mathbf{P}_\perp)}{\sqrt{P^+}},
\end{equation}
in which the light-cone spinors represent the electrons and
$\epsilon^\mu$ represents the polarization of the photon.

The same issue can be also analyzed in the light-cone quark model.
In this model, the light-cone wave function of a composite system
can be obtained by transforming the ordinary equal-time
(instant-form) wave function in the rest frame into that in the
light-front dynamics, by taking into account relativistic effects
such as the Melosh-wigner rotation. The equivalence of two
approaches has been demonstrated in Ref.~\cite{ref:xm}, where both
approaches are used to derive the pion light-cone wave function
and lead to the same result. The Melosh-Wigner rotation is one of
the most important ingredients of the light-cone formalism, and it
relates the light-cone spin state $|J,\lambda\rangle_F$ to the
ordinary instant-form spin state wave functions $|J,s\rangle_T$ by
the general relation~\cite{ref:wigner,ref:melosh,ref:ma1}
\begin{equation}
|J,\lambda\rangle_F=\sum_s
U^J_{s\lambda}|J,s\rangle_T.\label{eq:mw}
\end{equation}
The effects due to the Melosh-Wigner rotation have been calculated
for the nucleon axial charges~\cite{ref:ma1,bs94}, the magnetic
moments~\cite{bs94}, the nucleon helicity \cite{ma} and
transversity \cite{transversity} distributions, the quark orbital
angular moments of the nucleon \cite{LCor}, and recently for the
form factors of nucleons~\cite{ref:ma4}.

We now derive the light-cone wave function for the quark-diquark
Fock state of the proton in the light-cone quark model, with both
scalar diquark and vector diquark.
 The proton wave function in the SU(6) quark-diquark model in the instant form is written
 as~\cite{ma,ref:qdq}
 \begin{equation}
 \Psi^{\Uparrow,\Downarrow}(qD)=\textmd{sin} \theta \varphi_V|qV\rangle^{\Uparrow,\Downarrow}
 +\textmd{cos} \theta \varphi_S|qS\rangle^{\Uparrow,\Downarrow},\label{eq:qdq}
\end{equation}
with
\begin{eqnarray}
|qV\rangle^{\Uparrow,\Downarrow}&=&\pm\frac{1}{3}
[V_0(ud)u^{\uparrow,\downarrow}-
\sqrt{2}V_{\pm1}(ud)u^{\downarrow,\uparrow}-\sqrt{2}V_0(uu)d^{\uparrow,\downarrow}\nonumber\\
 &&+
2V_{\pm1}(uu)d^{\downarrow,\uparrow}];\nonumber\\
|qS\rangle^{\Uparrow,\Downarrow}&=&S(ud)u^{\uparrow,\downarrow},
\end{eqnarray}
where $\Uparrow,\Downarrow$ and $\uparrow,\downarrow$ label the
spin projection $J_p^z$=$\pm\frac{1}{2}$ of the proton and
$J_q^z$=$\pm\frac{1}{2}$ of the quark, respectively,
$V_{s_z}(q_1q_2)$ stands for the $q_1q_2$ vector diquark Fock
state with third spin component $s_z$, $S(ud)$ stands for a $ud$
scalar diquark Fock state, and $\varphi_D$ represents the momentum
space wave function of the quark-diquark state with $D$ denoting
the vector or scalar diquarks and $\theta$ being a mixing angle
that breaks SU(6) symmetry at $\theta\neq\pi/4$. Here we choose
$\theta=\pi/4$ following Ref.~\cite{ma}. Applying the
transformation Eq.~(\ref{eq:mw}) on both sides of
Eq.~(\ref{eq:qdq}), we will obtain the spin space wave function of
the quark-diquark Fock state in light-front frame. The proton
state with zero transverse momentum
$\mathbf{P}_\perp=\mathbf{0}_\perp$ in the \textit{l.h.s.} of
Eq.~(\ref{eq:qdq}) keeps identical under the Melosh-Wigner
rotation:
\begin{equation}
\Psi_F^{\Uparrow,\Downarrow}(P^+,P^-,\mathbf{0}_\perp)=\Psi_T^{\Uparrow,\Downarrow}(P^+,P^-,\mathbf{0}_\perp).
\end{equation}
For the \textit{r.h.s} of  Eq.~(\ref{eq:qdq}), we first consider
the spin part of quark-scalar-diquark Fock state,which is
determined by the spin of the quark $\chi_q$:
\begin{equation}
 \chi_{|qS\rangle}^{\Uparrow,\Downarrow}=\chi_q^{\uparrow,\downarrow}.
\end{equation}
The Melosh transformation on a quark spin state with four-momentum
$(k^0,\mathbf{k})$ is
\begin{eqnarray}
 \chi_{q}^\uparrow(T)&=&\omega_q[(k^++m)\chi_q^\uparrow(F)-k^R\chi_q^\downarrow(F)],\nonumber\\
 \chi_{q}^\downarrow(T)&=&\omega_q[(k^++m)\chi_q^\downarrow(F)+k^L\chi_q^\uparrow(F)],\label{eq:qr}
\end{eqnarray}
where $\omega_q=[2k^+(k^0+m)]^{-\frac{1}{2}}$,
$k^{R,L}=\mathbf{k}_\perp^1\pm \textmd{i}\mathbf{k}_\perp^2$, and
$k^+=k^0+k^3=x\mathcal{M}$, $x$ is the momentum fraction of the
quark, $m$ is the mass of the constituent quark, and $\mathcal{M}$
is the invariance mass of the composite state, which is
approximately taken as the proton mass $M$ in this work. Therefore
we get the light-cone spin wave function of the
quark-scalar-diquark part for the $J_p^z=+\frac{1}{2}$ proton:
\begin{equation}
\chi_{|qS\rangle}(x,\mathbf{k}_\perp)=\sum_{J^z_q}
C_{+\frac{1}{2}}^F(x,\mathbf{k}_\perp,J^z_q)\chi^{J^z_q},
\end{equation}
where the component coefficients
$C_{+\frac{1}{2}}^F(x,\mathbf{k}_\perp,J^z_q)$ have the forms:
\begin{eqnarray}
C_{+\frac{1}{2}}^F(x,\mathbf{k}_\perp,\uparrow)&=&\omega_q(xM+m),\nonumber\\
C_{+\frac{1}{2}}^F(x,\mathbf{k}_\perp,\downarrow)&=&-\omega_q
(\mathbf{k}_\perp^1+\textmd{i}\mathbf{k}_\perp^2).\label{eq:cfs}
\end{eqnarray}

With the momentum space wave function for the diquark, we can
write down the light-cone wave function for quark-scalar-diquark
Fock state as follows
\begin{eqnarray}
|Sq(P^+,\mathbf{P}_\perp&=&\mathbf{0}_\perp)\rangle^\Uparrow=\int
\frac{\textmd{d}^2\mathbf{k}_\perp
\textmd{d}x}{\sqrt{x(1-x)}16\pi^3}[\psi^\Uparrow_S(x,\mathbf{k}_\perp,\uparrow)
|xP^+,\mathbf{k}_\perp,\uparrow\rangle\nonumber\\
&&+\psi^\Uparrow_S(x,\mathbf{k}_\perp,\downarrow)
|xP^+,\mathbf{k}_\perp,\downarrow\rangle],\label{eq:sswf}
\end{eqnarray}
where
\begin{eqnarray}
\psi^\Uparrow_S(x,\mathbf{k}_\perp,\uparrow)&=&(M+\frac{m}{x})\varphi_S,\\
\psi^\Uparrow_S(x,\mathbf{k}_\perp,\downarrow)&=&-\frac{\mathbf{k}_\perp^1+\textmd{i}\mathbf{k}_\perp^2}{x}\varphi_S,
\end{eqnarray}
which is consistent with the light-cone wave function employed in
Ref.~\cite{ref:bhs}. For the momentum space wave function of the
scalar diquark we adopt
\begin{equation}
\varphi_S=\frac{e/\sqrt{(1-x)}}{M^2-(\mathbf{k}_\perp^2+m^2)/x-(\mathbf{k}_\perp^2+{\lambda_S}^2)/(1-x)},\label{eq:smswf}
\end{equation}
where $\lambda_S$ is the mass of the scalar diquark. The
$\omega_q$ appearing in Eq.~(\ref{eq:cfs}) can be considered to be
incorporated into the momentum space wave function $\varphi_S$. It
is shown that there is a component
$\psi^\Uparrow_S(x,\mathbf{k}_\perp,\downarrow)$ in the proton
wave function coming from the Wigner-Melosh rotation effect, which
does not exist in the non-relativistic constituent quark model.
Note that each spin configuration satisfies the spin sum rule
$J^z_q+L^z=+\frac{1}{2}$, where $L^z$ is the orbital angular
momentum projection.

Now consider the quark-vector-diquark two-body Fock state in the
right hand side of Eq.~(\ref{eq:qdq}). The spin part of the state
composed by $up$ quark and vector diquark with
$J^z_p=+\frac{1}{2}$ is written as
\begin{equation}
\chi_V^0\chi_u^\uparrow-\sqrt{2}\chi_V^{+1}\chi_u^\downarrow,\label{eq:swf}
\end{equation}
where $\chi_V$ stands for the spin of the vector diquark. For the
spin-1 vector diquark, the Melosh transformations
read~\cite{ref:as}
\begin{eqnarray}
\chi_V^{+1}(T)&=&\omega^2_V[(k_V^++\lambda_V)^2\chi^{+1}_V(F)-\sqrt{2}(k_V^++\lambda_V)k_V^R\chi^0_V(F)
+(k_V^R)^2\chi_V^{-1}(F)],\nonumber\\
\chi_V^{0}(T)&=&\omega^2_V\{\sqrt{2}(k_V^++\lambda_V)k_V^L\chi^{+1}_V(F)+2[(k_V^++\lambda_V)k_V^+-k_V^Rk_V^L]\chi^0_V(F)\nonumber\\
&&-\sqrt{2}(k_V^++\lambda_V)k_V^R\chi_V^{-1}(F)\},\nonumber\\
\chi_V^{-1}(T)&=&\omega^2_V[(k_V^L)^2\chi^{+1}_V(F)+\sqrt{2}(k_V^++\lambda_V)k_V^L\chi^0_V(F)
+(k_V^++\lambda_V)^2\chi_V^{-1}(F)]\label{eq:mwv},
\end{eqnarray}
where $\lambda_V$ is the mass of the vector diquark,
$k_V^+=k_{V}^0+k_V^3$=$(1-x)\mathcal{M}$,
$k_V^{R,L}=\mathbf{k}_{V\perp}^1\pm
\textmd{i}\mathbf{k}_{V\perp}^2$, here $\mathbf{k}_{V\perp}$ is
the transverse momentum of the vector diquark satisfied
$\mathbf{k}_{V\perp}=-\mathbf{k}_\perp$, and
$w_V=[2k_V^+(k_V^0+\lambda_V)]$.

Now substituting Eq.~(\ref{eq:mwv}) and Eq.~(\ref{eq:qr}) into
Eq.~(\ref{eq:swf}), we obtain the spin space wave function of the
$up$-quark-vector-diquark state for the $J^z_p=+\frac{1}{2}$
proton in the light-front frame:
\begin{equation}
\chi_{|Vu\rangle}(x,\mathbf{k}_\perp)=\sum_{J^z_{\small
V},J^z_q}C_{+\frac{1}{2}}^F(x,\mathbf{k}_\perp,J^z_{\small
V},J^z_q)\chi_V^{J^z_{\small V}}\chi_u^{J^z_q},
\end{equation}
where the coefficients read
\begin{eqnarray}
&&C_{+\frac{1}{2}}^F(x,\mathbf{k}_\perp,+1,\uparrow)=w_qw^2_V[-\sqrt{2}(k_V^++\lambda_V)(k^++m)-\sqrt{2}(k_V^++\lambda_V)^2]k^L,\nonumber\\
&&C_{+\frac{1}{2}}^F(x,\mathbf{k}_\perp,+1,\downarrow)=w_qw^2_V[\sqrt{2}(k_V^++\lambda_V)\mathbf{k}^2_\perp-\sqrt{2}(k_V^++\lambda_V)^2(k^++m)],\nonumber\\
&&C_{+\frac{1}{2}}^F(x,\mathbf{k}_\perp,0,\uparrow)=w_qw^2_V\{2[(k_V^++\lambda_V)k_V^+-\mathbf{k}^2_\perp](k^++m)-2(k_V^++\lambda_V)\mathbf{k}^2_\perp\},\nonumber\\
&&C_{+\frac{1}{2}}^F(x,\mathbf{k}_\perp,0,\downarrow)=w_qw^2_V2[-(k_V^++\lambda_V)k_V^++\mathbf{k}^2_\perp-2(k_V^++\lambda_V)(k^++m)]k^R,\nonumber\\
&&C_{+\frac{1}{2}}^F(x,\mathbf{k}_\perp,-1,\uparrow)=w_qw^2_V[\sqrt{2}(k_V^++\lambda_V)(k^++m)-\sqrt{2} \mathbf{k}^2_\perp]k^R,\nonumber\\
&&C_{+\frac{1}{2}}^F(x,\mathbf{k}_\perp,-1,\downarrow)=w_qw^2_V[\sqrt{2}
k^{R2}_\perp(k_V^++\lambda_V+k^++m)].\label{eq:vec}
\end{eqnarray}

Similar to the quark-scalar-diquark two-body Fock state wave
function, we notice that there are also higher helicity components
in the case of vector diquark. They also appear in the analysis of
the form factor of the nucleon~\cite{ref:ma4}. With $k^{R/L}$ in
the wave function, those transverse momentum dependent higher
helicity components $(+1,\downarrow)$, $(0,\downarrow)$,
$(-1,\uparrow)$ and $(-1,\downarrow)$ come from the Melosh-Wigner
rotation and reflect the intrinsic transverse motion of quarks in
the proton. Each spin configuration satisfies the spin sum rule
$J^z_q+J^z_{\small V}+L^z=+\frac{1}{2}$. The component
$(-1,\downarrow)$ is the higher twist component which can be
ignored in contrast with other components. From Eq.~(\ref{eq:vec})
we write down the expression for the light-cone wave function of
$up$-quark-vector-diquark Fock state with $J^z_p=+\frac{1}{2}$:
\begin{eqnarray}
&&|Vu(P^+,\mathbf{P}_\perp=\mathbf{0}_\perp)\rangle^\Uparrow=\int\frac{\textmd{d}^2\mathbf{k}_\perp
\textmd{d}x}{\sqrt{x(1-x)}16\pi^3}[\psi^\Uparrow_V(x,\mathbf{k}_\perp,+1,\uparrow)|xP^+,\mathbf{k}_\perp,+1,\uparrow\rangle
\nonumber\\&&+\psi^\Uparrow_V(x,\mathbf{k}_\perp,+1,\downarrow)|xP^+,\mathbf{k}_\perp,+1,\downarrow\rangle
+\psi^\Uparrow_V(x,\mathbf{k}_\perp,-1,\uparrow)|xP^+,\mathbf{k}_\perp,-1,\uparrow\rangle
\nonumber\\&&+\psi^\Uparrow_V(x,\mathbf{k}_\perp,0,\uparrow)|xP^+,\mathbf{k}_\perp,0,\uparrow\rangle
+\psi^\Uparrow_V(x,\mathbf{k}_\perp,0,\downarrow)|xP^+,\mathbf{k}_\perp,0,\downarrow\rangle],\label{eq:svwf}
\end{eqnarray}
in which
\begin{eqnarray}
&\psi^\Uparrow_V(x,\mathbf{k}_\perp,+1,\uparrow)&= -\sqrt{2}\frac{\mathbf{k}_\perp^1-\textmd{i}\mathbf{k}_\perp^2}{x(1-x)}\varphi_V,\nonumber\\
&\psi^\Uparrow_V(x,\mathbf{k}_\perp,+1,\downarrow)&= \sqrt{2}(M+\frac{\lambda_V}{1-x})\varphi_V,\nonumber\\
&\psi^\Uparrow_V(x,\mathbf{k}_\perp,-1,\uparrow)&= \sqrt{2}\frac{\mathbf{k}_\perp^1+\textmd{i}\mathbf{k}_\perp^2}{1-x}\varphi_V,\nonumber\\
&\psi^\Uparrow_V(x,\mathbf{k}_\perp,0,\uparrow)&= \Big{[}2(M+\frac{\lambda_V}{1-x})-(M+\frac{m}{x})\Big{]}\varphi_V,\nonumber\\
&\psi^\Uparrow_V(x,\mathbf{k}_\perp,0,\downarrow)&=
-\frac{1+x}{1-x}\frac{\mathbf{k}_\perp^1+\textmd{i}\mathbf{k}_\perp^2}{x}\varphi_V,\label{eq:vwf}
\end{eqnarray}
where $\varphi_V$ stands for the momentum space wave function,
which has the same form of Eq.~(\ref{eq:smswf}), but $\lambda_S$
replaced by $\lambda_V$. The above components of the wave function
are similar to the case of the light-cone wave function of
$|e\gamma\rangle$ Fock state in Ref.~\cite{BD80,ref:bhms}, except
that there are two components with spin projection $J^z_{\small
V}$=0, according to the mass of the vector diquark. We should
point out that Eq.~(\ref{eq:vwf}) is the parametrization of the
spin coupling of the quark-vector-diquark state, in which the
three spin projections of the vector diquark have been specified
explicitly. This is different from the spectator model employed in
Ref.~\cite{ref:bsy}, where the spin coupling is parameterized by a
covariant nucleon-quark-diquark interaction vertex.

Similarly we can obtain the Fock state of
$down$-quark-vector-diquark which is as the same as
Eq.~(\ref{eq:vwf}) only with an extra factor $-\sqrt{2}$, as shown
in following schematic form of the proton light-cone wave function
with $J^z_p=+\frac{1}{2}$:
\begin{equation}
\Psi_F^\Uparrow=\frac{1}{\sqrt{2}}|Su\rangle_F^\Uparrow+\frac{1}{3\sqrt{2}}|Vu\rangle_F^\Uparrow-\frac{1}{3}|Vd\rangle_F^\Uparrow.\label{eq:lcwf}
\end{equation}
The $J^z_p=-\frac{1}{2}$ proton wave function can be evaluated in
the same way.

\section{Estimates of Sivers function for up and down quark}

The final state interactions are generated by the gauge link term
in the gauge-invariance definition of the parton distribution
funcitons. In model
calculations~\cite{ref:bsy,ref:bbh,ref:yuan,ref:gamberg} of
non-vanishing Sivers function, one gluon-exchange approximation
for the gauge link exponential was adopted. In the language of
final state interactions, the nonzero phase comes from the
interference of proton spin amplitudes, in which different proton
spin states $J^z_p=\pm\frac{1}{2}$ couple to the same final state
$|F\rangle$, requiring the orbital angular momentum of the two
proton wave functions differing by $\Delta L^z=1$. In previous
section, we have derived the light-cone wave function including
quark-vector-diquark structure. Like the quark-scalar-diquark
state in Ref.~\cite{ref:bhs}, the quark-vector-diquark structure
also contributes to the proton spin amplitudes, and further the
final state interactions. Following the calculation in
Ref.~\cite{ref:bhs}, we calculate all possible proton spin
amplitudes, by adopting one gluon-exchange approximation:
\begin{eqnarray}
&&\mathcal{A}(\Uparrow\rightarrow -1\uparrow
)=\sqrt{2}\frac{\mathbf{r}_\perp^1+\textmd{i}\mathbf{r}_\perp^2}
{1-\Delta}C(h_v+\textmd{i}\frac{e_1e_2}{8\pi}g_{v2}),\nonumber\\
&&\mathcal{A}(\Uparrow\rightarrow +1\downarrow
)=-\sqrt{2}(M+\frac{\lambda_V}{1-\Delta})C(h_v+\textmd{i}\frac{e_1e_2}{8\pi}g_{v1}),\nonumber\\
&&\mathcal{A}(\Uparrow\rightarrow +1\uparrow
)=-\sqrt{2}\frac{\mathbf{r}_\perp^1-\textmd{i}\mathbf{r}_\perp^2}{\Delta(1-\Delta)}C(h_v+\textmd{i}\frac{e_1e_2}{8\pi}g_{v1}),\nonumber\\
&&\mathcal{A}(\Uparrow\rightarrow 0\uparrow
)=(M+2\frac{\lambda_V}{1-\Delta}-\frac{m}{\Delta})C(h_v+\textmd{i}\frac{e_1e_2}{8\pi}g_{v1}),\nonumber\\
&&\mathcal{A}(\Uparrow\rightarrow 0\downarrow
)=-\frac{1+\Delta}{1-\Delta}C\frac{\mathbf{r}_\perp^1+\textmd{i}\mathbf{r}_\perp^2}{\Delta}(h_v+\textmd{i}\frac{e_1e_2}{8\pi}g_{v1}),\nonumber\\
&&\mathcal{A}(\Uparrow\rightarrow S\uparrow
)=(M+\frac{m}{\Delta})C(h_s+\textmd{i}\frac{e_1e_2}{8\pi}g_{s1}),\nonumber\\
&&\mathcal{A}(\Uparrow\rightarrow S\downarrow
)=-\frac{\mathbf{r}_\perp^1+\textmd{i}\mathbf{r}_\perp^2}{\Delta}C(h_s+\textmd{i}\frac{e_1e_2}{8\pi}g_{s2}),\label{eq:protonup}
\end{eqnarray}
for $J^z_p=+\frac{1}{2}$ two-body states and
\begin{eqnarray}
&&\mathcal{A}(\Downarrow\rightarrow -1\uparrow
)=-\sqrt{2}(M+\frac{\lambda_V}{1-\Delta})C(h_v+i\frac{e_1e_2}{8\pi}g_{v1}),\nonumber\\
&&\mathcal{A}(\Downarrow\rightarrow +1\downarrow
)=-\sqrt{2}\frac{\mathbf{r}_\perp^1-\textmd{i}\mathbf{r}_\perp^2}{1-\Delta}C(h_v+\textmd{i}\frac{e_1e_2}{8\pi}g_{v2}),\nonumber\\
&&\mathcal{A}(\Downarrow\rightarrow -1\downarrow
)=\sqrt{2}\frac{\mathbf{r}_\perp^1+\textmd{i}\mathbf{r}_\perp^2}{\Delta(1-\Delta)}C(h_v+\textmd{i}\frac{e_1e_2}{8\pi}g_{v2}),\nonumber\\
&&\mathcal{A}(\Downarrow\rightarrow 0\uparrow
)=\frac{1+\Delta}{1-\Delta}(\mathbf{r}_\perp^1-\textmd{i}\mathbf{r}_\perp^2)C(h_v+\textmd{i}\frac{e_1e_2}{8\pi}g_{v2}),\nonumber\\
&&\mathcal{A}(\Downarrow\rightarrow 0\downarrow
)=(M+2\frac{\lambda_V}{1-\Delta}-\frac{m}{\Delta})C(h_v+\textmd{i}\frac{e_1e_2}{8\pi}g_{v2}),\nonumber\\
&&\mathcal{A}(\Downarrow\rightarrow S\uparrow
)=\frac{\mathbf{r}_\perp^1-i\mathbf{r}_\perp^2}{\Delta}C(h_s+\textmd{i}\frac{e_1e_2}{8\pi}g_{s2}),\nonumber\\
&&\mathcal{A}(\Downarrow\rightarrow S\downarrow
)=(M+\frac{m}{\Delta})C(h_s+\textmd{i}\frac{e_1e_2}{8\pi}g_{s1}),\label{eq:protondown}
\end{eqnarray}
for $J^z_p=-\frac{1}{2}$ two-body states. Where $\Delta$ and
$\mathbf{r}_\perp$ are the Bjorken variable and the transverse
momentum of the struck quark respectively, $e_1$ and $e_2$ are the
charges of the struck quark and spectator diquark, and
\begin{eqnarray}
&&C=-g e_1P^+\sqrt{\Delta}2\Delta(1-\Delta),\\
&&h_{s/v}=\frac{1}{\mathbf{r}_\perp+\Delta(1-\Delta)(-M^2+\frac{m^2}{\Delta}+\frac{\lambda_{S/V}}{1-\Delta})},
\end{eqnarray}
in which $g$ is the proton-quark-diquark coupling constant. In
literature there are two choice for $g$, one is to take $g$ as a
constant~\cite{ref:bhs,ref:bbh}, another as a form
factor~\cite{ref:bsy,ref:gamberg} relying on the momentum of the
outgoing quark, to converge the transverse momentum integration of
the distribution functions. As argued in Ref.~\cite{ref:bbh}, the
inclusion of the form factor would add another complication on the
evolution of the asymmetry, therefore we treat $g$ as a constant
in our model, which is equal for the scalar and the vector
diquark. The last two equations in Eq.~(\ref{eq:protonup}) and
Eq.~(\ref{eq:protondown}) are as the same as Eq.~(5)-Eq.~(8) in
Ref.~\cite{ref:bhs}, while the other ten equations are contributed
from vector-diquark structure. With above amplitudes we calculate
the asymmetries in quark level for $up$ and $down$ quarks
through~\cite{ref:bhs2}
\begin{equation}
\mathcal{P}_y^{u/d}=\frac{\textmd{Im}\Big{(}\mathcal{T}_{int}^{u/d}\Big{)}}{\mathcal{M}^{u/d}}.\label{eq:py}
\end{equation}
Here $\mathcal{T}_{int}^u$ and $\mathcal{T}_{int}^d$  are the
interference terms coming from the final state interactions with
the forms
\begin{eqnarray}
\mathcal{T}_{int}^u&=&\mathcal{A}(\Uparrow\rightarrow
S\uparrow)^*\mathcal{A}(\Downarrow\rightarrow S\uparrow)
+\frac{1}{9}\mathcal{A}(\Uparrow\rightarrow
-1\uparrow)^*\mathcal{A}(\Downarrow\rightarrow
-1\uparrow)\nonumber\\&&+\frac{1}{9}\mathcal{A}(\Uparrow\rightarrow
0\uparrow)^*\mathcal{A}(\Downarrow\rightarrow 0\uparrow),\nonumber\\
\mathcal{T}_{int}^d&=&\frac{2}{9}\mathcal{A}(\Uparrow\rightarrow
-1\uparrow )^*\mathcal{A}(\Downarrow\rightarrow -1\uparrow)
+\frac{2}{9}\mathcal{A}(\Uparrow\rightarrow
0\uparrow)^*\mathcal{A}(\Downarrow\rightarrow 0\uparrow),
\end{eqnarray}
respectively, and $\mathcal{M}^{u/d}$ are denoted as:
\begin{eqnarray}
\mathcal{M}^u&=&\frac{1}{2}\Big{|}\mathcal{A}(\Uparrow\rightarrow
S\uparrow
)\Big{|}^2+\frac{1}{18}\Big{|}\mathcal{A}(\Uparrow\rightarrow
S\downarrow
)\Big{|}^2+\frac{1}{18}\Big{|}\mathcal{A}(\Uparrow\rightarrow
+1\uparrow)\Big{|}^2\nonumber\\&&+\frac{1}{18}\Big{|}\mathcal{A}(\Uparrow\rightarrow
+1\downarrow)\Big{|}^2+
\frac{1}{18}\Big{|}\mathcal{A}(\Uparrow\rightarrow -1\downarrow
)\Big{|}^2+\frac{1}{18}\Big{|}\mathcal{A}(\Uparrow\rightarrow
0\uparrow
)\Big{|}^2\nonumber\\&&+\frac{1}{18}\Big{|}\mathcal{A}(\Uparrow\rightarrow
0\downarrow)\Big{|}^2,
\nonumber\\
\mathcal{M}^d&=& \frac{1}{9}\Big{|}\mathcal{A}(\Uparrow\rightarrow
+1\uparrow)\Big{|}^2+\frac{1}{9}\Big{|}\mathcal{A}(\Uparrow\rightarrow
+1\downarrow)\Big{|}^2+\frac{1}{9}\Big{|}\mathcal{A}(\Uparrow\rightarrow
-1\downarrow)\Big{|}^2\nonumber\\&
&+\frac{1}{9}\Big{|}\mathcal{A}(\Uparrow\rightarrow
0\uparrow)\Big{|}^2+\frac{1}{9}\Big{|}\mathcal{A}(\Uparrow\rightarrow
0\downarrow)\Big{|}^2.
\end{eqnarray}

Substituting Eq.~(\ref{eq:protonup}) and Eq.~(\ref{eq:protondown})
into the equations above we obtain
\begin{eqnarray}
\mathcal{T}_{int}^u&=&-\frac{1}{2}(M+\frac{m}{\Delta})\frac{\mathbf{r}_\perp^1}{\Delta}
\frac{e_1e_2}{4\pi}\frac{1}{\Lambda_s(\mathbf{r}^2_\perp)\mathbf{r}_\perp^2}
\textmd{ln}\frac{\Lambda_s(\mathbf{r}^2_\perp)}{\Lambda_s(0)}\nonumber\\
&&-\Big{[}\frac{2}{9}(M+\frac{\lambda_V}{1-\Delta})\frac{\mathbf{r}_\perp^1}{1-\Delta}
+\frac{1}{9}(M+2\frac{\lambda_V}{1-\Delta}-\frac{m}{\Delta})\nonumber\\
&&\cdot\frac{1-\Delta}
{(1+\Delta)}\frac{\mathbf{r}_\perp^1}{\Delta}\Big{]}\frac{e_1e_2}{8\pi}\frac{1}{\Lambda_v(\mathbf{r}^2_\perp)\mathbf{r}_\perp^2}
\textmd{ln}\frac{\Lambda_v(\mathbf{r}^2_\perp)}{\Lambda_v(0)},\\
\mathcal{T}_{int}^d&=&-\frac{e_1e_2}{4\pi}\frac{1}{\Lambda_v(\mathbf{r}^2_\perp)
\mathbf{r}^2_\perp}\textmd{ln}\frac{\Lambda_v(\mathbf{r}^2_\perp)}{\Lambda_v(0)}
\Big{[}\frac{2}{9}(M+\frac{\lambda_V}{1-\Delta})\frac{\mathbf{r}_\perp^1}{1-\Delta}\nonumber\\&
&+\frac{1}{9}(M+2\frac{\lambda_V}{1-\Delta}-\frac{m}{\Delta})\frac{1-\Delta}
{(1+\Delta)}\frac{\mathbf{r}_\perp^1}{\Delta}\Big{]},\\
\mathcal{M}^u&=&\frac{1}{2}\Big{[}(M+\frac{m}{\Delta})^2+\frac{\mathbf{r}_\perp^2}
{\Delta^2}\Big{]}h_s^2+\frac{1}{9}\Big{[}(M+\frac{\lambda_V}{1-\Delta})^2
\frac{\mathbf{r}_\perp^2}{(1-\Delta)^2} \nonumber\\&
&+\frac{\mathbf{r}_\perp^2}{\Delta^2(1-\Delta)^2}
+\frac{(1-\Delta)^2}{(1+\Delta)^2}\frac{\mathbf{r}_\perp^2}{2\Delta^2}+(\frac{1}{2}M+\frac{\lambda_V}{1-\Delta}-\frac{m}{2\Delta})^2\bigg{]}h_v^2,\\
\mathcal{M}^d&=&\frac{2}{9}\Big{[}(M+\frac{\lambda_V}{1-\Delta})^2+\frac{\mathbf{r}_\perp^2}
{(1-\Delta)^2}+\frac{\mathbf{r}_\perp^2}{\Delta^2(1-\Delta)^2}
+\frac{(1-\Delta)^2}{(1+\Delta)^2}\frac{\mathbf{r}_\perp^2}{2\Delta^2}\nonumber\\&
&+(\frac{1}{2}M+\frac{\lambda_V}{1-\Delta}-\frac{m}{2\Delta})^2\bigg{]}h_v^2,
\end{eqnarray}
in which
\begin{equation}
\Lambda_{s/v}(\mathbf{r}^2_\perp)=\mathbf{r}^2_\perp+\Delta(1-\Delta)(-M^2+\frac{m}{\Delta}+\frac{\lambda_{S/V}}{1-\Delta}).
\end{equation}

From Eq.~(\ref{eq:py}) and the relation between the asymmetries
$\mathcal{P}_y$, $f^\perp_{1T}(\Delta,\mathbf{r}_\perp^2)$ and
$f_1(\Delta,\mathbf{r}_\perp^2)$~\cite{ref:bhs2,ref:bbh}
\begin{equation}
\mathcal{P}_y^a=-\frac{\mathbf{r}_\perp^1}{M}f^{\perp
a}_{1T}(\Delta,\mathbf{r}_\perp^2)/f^a_1(\Delta,\mathbf{r}_\perp^2),\label{eq:relation}
\end{equation}
we obtain the Sivers functions of $up$ and $down$ quarks by
\begin{equation}
f^{\perp
a}_{1T}(\Delta,\mathbf{r}_\perp^2)=-\frac{\mathcal{P}_y^a}{\mathbf{r}_\perp^1}f_1^a(\Delta,\mathbf{r}_\perp^2)\label{pya}.
\end{equation}
Here $a$ is the flavor index including $u$ and $d$. The difference
between Eq.~(\ref{eq:relation}) and Eq.~(32) in
Ref.~\cite{ref:bbh} is that we assume that the equation stands for
both $up$ and $down$ quarks. From Eq.~(\ref{eq:relation}) we can
see that if we exclude vector diquark, the result of our model
will return to the results in Ref.~\cite{ref:bhs} and
Ref.~\cite{ref:bbh}; when vector diquark is included, the Sivers
function for $down$ quark is evaluated. There is another advantage
of Eq.~(\ref{eq:py}): the coupling constant $g$ will appear in
both the numerator and denominator of Eq.~(\ref{eq:relation}) and
cancel, thus the Sivers functions in our model is free from $g$
and normalization factors. To avoid ambiguity, we use below the
same notation used in Ref. \cite{ref:bhs} with $\Delta$ denoting
the Bjorken variable of parton and $\mathbf{r}_\perp$ for the
transverse momentum of quark. In realistic application the moment
of the Sivers function is more useful. The
$\mathbf{r}^2_\perp$-moment of the Sivers function is defined
as~\cite{ref:tm}
\begin{equation}
f_{1T}^{\perp(1)}(\Delta)=\int d^2\mathbf{r}_\perp
\bigg{(}\frac{\mathbf{r}_\perp^2}{2M^2}\bigg{)}f_{1T}^\perp(\Delta,\mathbf{r}_\perp^2).\label{eq:moment}
\end{equation}

\begin{figure*}
\includegraphics{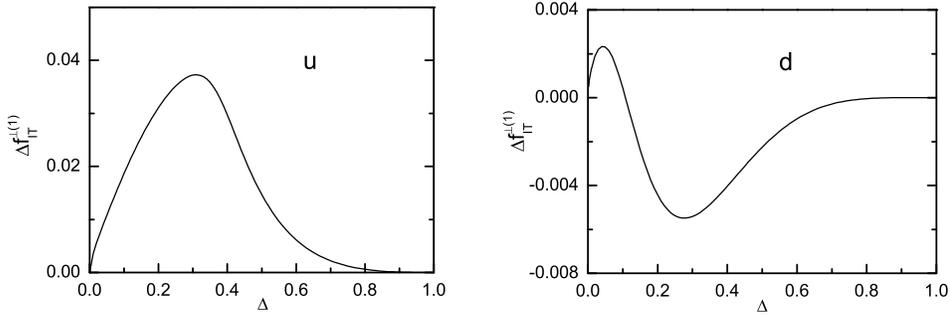}\caption{\small
The $\mathbf{r}_\perp^2$-moments of Sivers functions of $u$ and
$d$ quarks in SU(6) light-cone quark-diquark
model.}\label{fig:Sivfunction}
\end{figure*}

In numerical estimates of this moment we adopt Gaussian
distribution of the transverse momentum for the unpolarized
distribution function
\begin{equation}
f_{1}(\Delta,\mathbf{r}_\perp^2)=\frac{f_1(\Delta)}{\pi\langle
\mathbf{r}_\perp\rangle^2}{\textmd{Exp}\bigg{(}-\frac{\mathbf{r}^2_\perp}
{\langle\mathbf{r}_\perp\rangle^2}\bigg{)}},
\end{equation}
in order to guarantee convergence of the $\mathbf{r}_\perp$
integral in Eq.~(\ref{eq:moment}). Here
$\langle\mathbf{r}_\perp\rangle$ is the average transverse
momentum of the quark. For the transverse momentum integrated
unpolarized distribution $f_1(\Delta)$ we adopt CTEQ6M
parametrization~\cite{ref:cteq6} for the valence quarks. Also we
choose the parameters in the expression as: $\lambda_S=0.6$~GeV,
$\lambda_V=0.9$~GeV as estimated to explain the $N$-$\Lambda$ mass
difference,
 $m=0.36$~GeV, $M=0.94$~GeV,
$\langle\mathbf{r}_\perp\rangle=0.4$~GeV. To generalize the
analysis to the corresponding calculation in QCD, one needs to
extrapolate $|\frac{e_1e_2}{4\pi}|\rightarrow C_F\alpha_S$ with
$C_F=\frac{4}{3}$. We choose $\alpha_s=0.2$ to correspond the
kinematics of HERMES experiment. With above parameters we obtain
the Sivers functions of $up$ and $down$ quarks explicitly. The
$\mathbf{r}_\perp^2$-moments of the Sivers functions of $up$ and
$down$ quarks are figured by Fig.~\ref{fig:Sivfunction} which
shows that in our model the Sivers function of $up$ quark is
positive, while that of $down$ quark tends to be negative, and the
former is 6$\sim$7 times larger than the later in scale. This
agrees with the $\mathbf{r}_\perp$-moments of the Sivers functions
that appeared in Ref.~\cite{ref:bsy}. And the size of the moments
is similar to that of the bag-model result presented in
Ref.~\cite{ref:yuan}.

\section{Single-spin Sivers asymmetries of pion electroproduction in SIDIS}

\begin{figure*}
\begin{center}
\scalebox{0.8}{\includegraphics{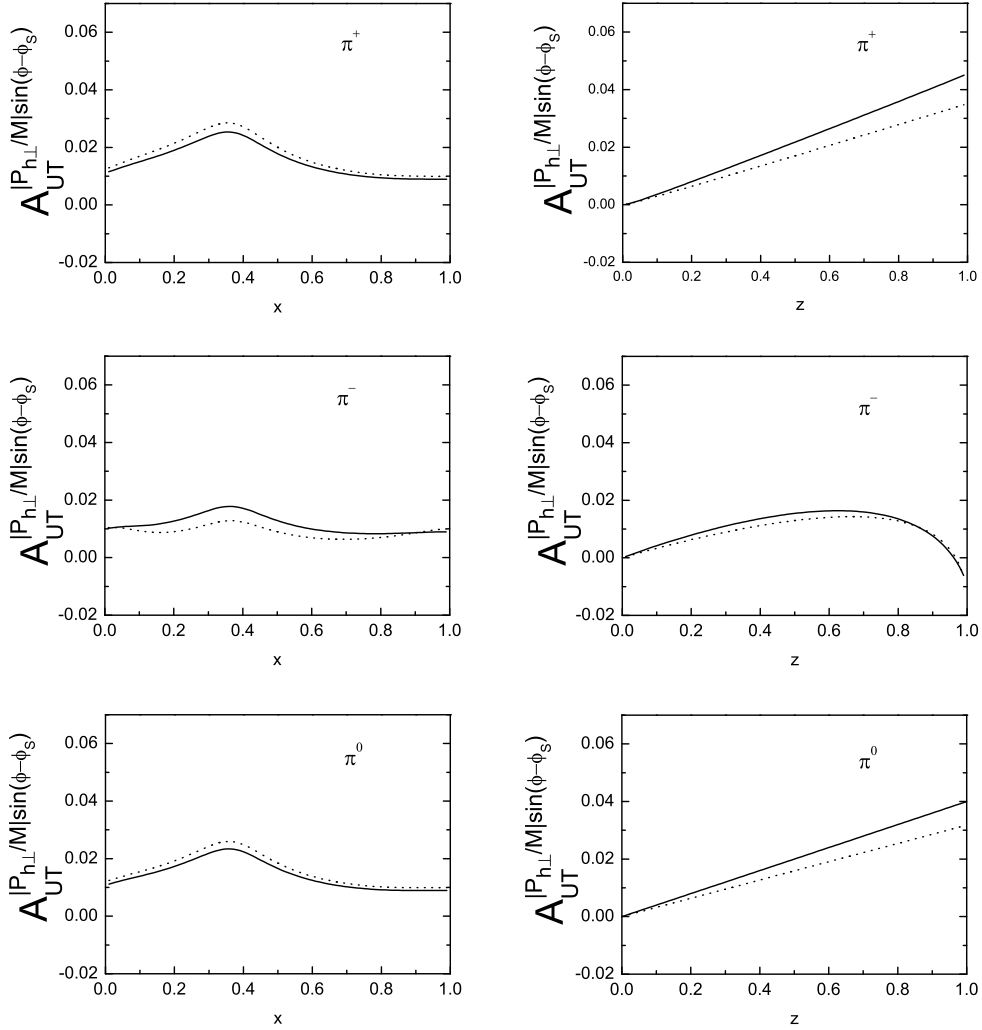}}
\caption{\small Single-spin Sivers asymmetries measured in SIDIS
off proton target. The asymmetries have been weighted by
$\frac{|P_{h\perp}|}{M}$. The first, second and third rows
correspond to the asymmetries of $\pi^+$, $\pi^-$ and $\pi^0$
production respectively. The left column corresponds to the
$x$-dependent asymmetries, the solid and dotted curves correspond
to the asymmetries averaged and not averaged by the HERMES
kinematics $0.2<z<0.7$. The right column corresponds to the
$z$-dependent asymmetries, the solid and dotted curves correspond
to the asymmetries averaged and not averaged by the HERMES
kinematics $0.023<x<0.4$. }\label{fig:Sivers_asymmetry}
\end{center}
\end{figure*}

Sivers effect can be observed in meson electroproduction in
semi-inclusive DIS with unpolarized electron beam off transversely
polarized nucleon target which is measured in HERMES. With the
Sivers functions of $up$ and $down$ quarks obtained above, we
calculate single-spin asymmetries in SIDIS contributed from Sivers
effect by selecting proper weighting factor
$\frac{|P_{h\perp}|}{M}\textmd{sin}(\phi-\phi_S)$. $\phi$ is the
azimuthal angle between the lepton scattering plane and the
transverse momentum of outgoing hadron, $\phi_S$ is the azimuthal
angle of the spin of proton target. The asymmetries are expressed
by the convolution of the $\mathbf{r}^2_\perp$-moment of Sivers
function and the usual unpolarized fragmentation function
$D^a_1(z)$, divided by the unpolarized cross section~\cite{ref:bm}
\begin{equation}
\bigg{\langle}\frac{|P_{h\perp}|}{M}\textmd{sin}(\phi-\phi_S)\bigg{\rangle}_{UT}(\Delta,z)
=\frac{\sum_ae^2_af^{\perp(1)a}_{1T}(\Delta)zD_1^a(z)
}{\sum_ae^2_af_{1}^{a}(\Delta)D_1^a(z)},
\end{equation}
where the subscript $UT$ represents unpolarized beam on
transversely polarized target, $P_{h\perp}$ is the transverse
momentum of the pion. For $D_1(z)$, we consider both the favored
fragmentation function
\begin{equation}
D(z)=D_u^{\pi^+}(z)=D_{d}^{\pi^-}(z)
\end{equation}
and unfavored fragmentation function~\cite{ref:msy}
\begin{equation}
\hat{D}(z)=D_d^{\pi^+}(z)=D_u^{\pi^-}(z)
\end{equation}
for $\pi^\pm$ fragmentation. Also we assume
\begin{equation}
D^{\pi^0}(z)=\frac{1}{2}[D_d^{\pi^+}(z)+D_u^{\pi^-}(z)]=\frac{1}{2}[D(z)+\hat{D}(z)]
\end{equation}
for $\pi^0$ fragmentation. For the explicit analytically forms of
$D(z)$ and $\hat{D}(z)$ we adopt~\cite{ref:kretzer}
\begin{eqnarray}
D(z)=0.689z^{-1.039}(1-z)^{1.241},\nonumber\\
\hat{D}(z)=0.217z^{-1.805}(1-z)^{2.037}.
\end{eqnarray}
For $\pi^-$ production, the favored fragmentation process is
$d\rightarrow\pi^-$. Since we have obtained the Sivers function of
$down$ quark, we can give some prediction of the azimuthal
asymmetries for the $\pi^-$ production.  We have ignored the sea
quark contribution, and assume that the contribution comes mainly
from the valence part, since we can not get information about sea
quark from the SU(6) quark-diquark model. For comparison with
future HERMES experimental data we also calculate the asymmetries
averaged by the kinematics of HERMES: $0.023<x<0.4$ and
$0.2<z<0.7$. The single-spin Sivers asymmetries of $\pi^+$,
$\pi^-$ and $\pi^0$ plotted vs $x$ and $z$ are presented in
Fig.~\ref{fig:Sivers_asymmetry}. The figure shows that the
$x$-dependent asymmetries of $\pi^+$ production are larger than
that of $\pi^-$ production. The $z$-dependent asymmetries of
$\pi^+$ and $\pi^0$ are similar, and different from that of
$\pi^-$ in shape and size.

\section{Conclusion}

Final state interactions provide a reliable explanation for Sivers
single-spin asymmetries in SIDIS. To get phenomenological
estimation we derive the light-cone wave function of the proton
based on the SU(6) quark-spectator-diquark model. We obtain the
light-cone wave function containing both scalar diquark and vector
diquark. The transverse momentum dependence, which is responsible
for the transverse momentum distribution function and single-spin
asymmetries, is introduced to the wave function by employing the
Wigner-Melosh rotation. The Wigner-Melosh rotation also brings the
higher helicity components which contribute to the final state
interactions. Utilizing the wave function we calculate the Sivers
functions of $up$ and $down$ quarks in one-gluon exchange. We find
that in our model the Sivers function of $up$ quark is positive,
while that of $down$ quark tends to be negative, and that the
former is much larger than the later in scale. We give analysis of
$\frac{|P_{h\perp}|}{M}$ weighted Sivers asymmetries
$A_{UT}^{\textmd{sin}(\phi-\phi_S)}$ of $\pi^+$, $\pi^-$ and
$\pi^0$ production off transverse proton targets plotted vs $x$
and $z$ from our model. The numerical result shows that the
$x$-dependent asymmetries of $\pi^+$ production are a little
larger than that of $\pi^-$ production. The $z$-dependent
asymmetries of $\pi^+$ and $\pi^0$ are similar, and different from
that of $\pi^-$ in shape and size.

In contrast to the result in Ref.~\cite{ref:bsy}, our Sivers
functions and asymmetries are qualitatively similar but
quantitatively different. We conclude that the differences come
from two aspects. On the one hand, we use different
quark-vector-diquark structure from that of Ref.~\cite{ref:bsy};
on the other hand, we treat proton-quark-diquark coupling $g$ as a
constant, different from that of Ref.~\cite{ref:bsy}. As a result,
the Sivers functions in our model are free from $g$ and
normalization parameters, but rely on $\langle
\mathbf{r}_\perp\rangle$.

Similar to other model calculations, we use some assumptions and
approximations, such as one Abelian gluon approximation and
valence quark dominance, and do not consider the evolution of the
Sivers function, thus at current level, our Sivers functions and
asymmetries are rough estimates.

{\bf Acknowledgements}

This work is partially supported by National Natural Science
Foundation of China under Grant Numbers 10025523 and 90103007.


\end{document}